\documentclass[times,10pt,twocolumn]{article}

\usepackage{epsfig}
\usepackage{subfigure}
\usepackage{calc}
\usepackage{amssymb}
\usepackage{amstext}
\usepackage{amsmath}
\usepackage{amsthm}
\usepackage{tabu,multicol}
\usepackage{multirow}
\usepackage{pslatex}
\usepackage{verbatim}
\usepackage{makecell}
\usepackage{caption}
\usepackage{SEKE}     % Please add other packages that you may need BEFORE the SCITEPRESS.sty package.

\begin{document}

\title{Logical Segmentation of Source Code}

\author{Jacob Dormuth, Ben Gelman, Jessica Moore, David Slater}
\affiliation{Machine Learning Group \\ Two Six Labs \\ Arlington, Virginia, United States}
\email{\{jacob.dormuth, ben.gelman, jessica.moore, david.slater\}@twosixlabs.com}

%\author{\authorname{Jacob Dormuth\sup{1}, Ben Gelman\sup{1}, Jessica Moore\sup{1}, David Slater\sup{1}}
%\affiliation{\sup{1}Two Six Labs, 4350 N. Fairfax Dr. Suite 410, Arlington, Virginia, United States of America}
%\email{jacob.dormuth@twosixlabs.com, ben.gelman@twosixlabs.com, jessica.moore@twosixlabs.com, david.slater@twosixlabs.com}
%}

%\keywords{The paper must have at least one keyword. The text must be set to 9-point font size and without the use of bold or italic font style. For more than one keyword, please use a comma as a separator. Keywords must be titlecased.}

\maketitle

%\onecolumn \maketitle \normalsize \vfill

\begin{abstract}

    Many software analysis methods have come to rely on machine learning approaches. Code segmentation - the process of decomposing source code into meaningful blocks - can augment these methods by featurizing code, reducing noise, and limiting the problem space. Traditionally, code segmentation has been done using syntactic cues; current approaches do not intentionally capture logical content. We develop a novel deep learning approach to generate logical code segments regardless of the language or syntactic correctness of the code. Due to the lack of logically segmented source code, we introduce a unique data set construction technique to approximate ground truth for logically segmented code. Logical code segmentation can improve tasks such as automatically commenting code, detecting software vulnerabilities, repairing bugs, labeling code functionality, and synthesizing new code. 

\end{abstract}

\section{\uppercase{Introduction}}
\label{sec:introduction}

\begin{figure}[b]
DOI reference number: 10.18293/SEKE2019-026 
\end{figure}

With the proliferation of open-source development practices and code sharing services, such as GitHub and Bitbucket, large bodies of source code are increasingly available to developers. There are a number of ways in which these code corpora could assist with the software development process; of particular interest is the application of machine learning to software engineering practices. Recent literature has utilized machine learning with source code at scale, developing tools to generate comments \cite{autocomment}, detect software vulnerabilities \cite{vulndetect}, repair bugs \cite{vulnrepair}, label functionality \cite{reversecrowd}, and synthesize new code \cite{musynth}. Code segmentation - the process of decomposing source code into meaningful blocks - can augment these methods, for instance, by determining what portions of a file are functionally similar, identifying where to generate automatic comments, and locating useful sub-function boundaries for bug detection.

Current approaches to segmentation do not intentionally capture logical content that could improve its usefulness to the aforementioned problems. Traditionally, code segmentation has been done at a syntactic level. This means that language-specific syntax, such as the closing curly brace of a class or function, is the indicator for a segment. Although this method is conceptually simple, the resulting segments do not intentionally take into account semantic information. Syntactic structures in natural language, such as sentences and paragraphs, are generally also indicators of semantic separation. In other words, two different paragraphs likely capture two logically separate ideas, and it is simple to delineate them by splitting at the end of each paragraph. Locating logical segments in source code, however, is a non-trivial task. Syntactic structures, such as a for-loop followed by an if-statement, are not particularly indicative of semantic changes in the code. Determining if the for-loop and if-statement are working to achieve the same logical task is a more difficult problem than the paragraph delineation analogue. Logical code segmentation captures arbitrary combinations of syntactic structures in a single segment.

Logical code segmentation has a multitude of uses, including: improving code search tools by returning relevant segments of code instead of entire files or functions, recommending locations to add comments, classifying the functionality of source code by reducing the problem space from entire files/projects to concentrated blocks of code, or using the segments as features for a model that attempts to determine the modularity or complexity of a given code file. Logical segments are able to featurize code, reduce noise, and limit a problem space to certain types of segments.

In this work, we develop a novel code segmentation method to generate logical segments in a language-agnostic fashion. Although language specificity may help model code by allowing options such as language-specific tokens and abstract syntax trees, it adds a non-trivial burden to operationalizing tools. Language specific approaches require training many models, each with their own parsers, training sets, and hyperparameters. Our language-agnostic approach is able to avoid those extra steps, improving the generalizability, availability, and ease-of-use of the results. Because there is no data set of source code that is conveniently split into logical segments, we first establish a unique data set construction process using Stack Overflow\footnote{https://stackoverflow.com} (SO). Human curation is critical to determining logical segments, and Stack Overflow provides that at a vast scale.

We build on prior work in natural language processing, training a bidirectional long short-term memory (LSTM) neural network to split source code into logical segments. Since this is the first work to perform this segmentation in the source code domain, we provide baseline models and multiple deep neural networks for comparison. We validate the results on six programming languages to display the language agnosticism of the method. Lastly, we qualitatively discuss the model's results on real source code documents to provide insight on performance in the desired domain. 

Our main contributions are as follows:
\begin{itemize}
  \item A novel data set construction method that utilizes crowd-sourced data to approximate logical segments. 
  \item First work, to our knowledge, to perform logical code segmentation regardless of syntactical correctness or programming language.
  \item Baseline models for the logical segmentation problem, demonstrating the relative effectiveness of our language-agnostic, deep neural network on six different programming languages. 
\end{itemize}

\section{\uppercase{Related Work}}
\label{sec:related}

The literature on source code segmentation is relatively sparse, with existing work offering syntax-specific solutions and requiring language specificity. Ning, Engberts, and Kozaczynski develop a code browser that allows users to select useful, reusable segments by analyzing the control flow and data flow in the abstract syntax trees (ASTs) of COBOL programs \cite{cobolsegment}. More recently, Wang, Pollock, and Vijay-Shanker attempt to locate logical segments by developing rules that look for specific syntactical patterns \cite{readability}. Wang et al. generate an AST for a given Java method, apply rules that analyze the data flow and syntactic structures, and then add line breaks around the resulting segments to enhance the readability of the method. Although these methods provide sophisticated analysis and rules, they are inextricably linked to the language that they operate on, and thus will not transfer to other languages.

In this work, we deviate from the practice of using ASTs as the foundation for our analysis. By treating source code as a body of text akin to an NLP problem, we avoid any programming-language-specific challenges posed by other methods. Text segmentation has been researched more thoroughly than the source code analogue, with methods ranging from LDA \cite{ldatextseg}, to semantic relatedness graphs \cite{graphtextseg}, to deep learning approaches \cite{textsegattention}. Of particular note is the use of bidirectional LSTMs to identify the breaks between segments of Wikipedia articles \cite{textsegdeeplstm}. An LSTM is a recurrent neural network that processes sequential information \cite{lstm}. Although we use the bidirectional LSTM as the primary framework of our model, we make several changes to adapt to the source code domain. We use a character embedding instead of the commonly used word embedding: due to the vast number of possible unique identifiers in source code, optimizing an embedding for each token is infeasible. Additionally, we require an evaluation metric specifically based on our data generation method to represent how well the model recognizes segments the method creates. 

\section{\uppercase{Data Set Generation}}
\label{sec:data}

\begin{figure}[t]
\includegraphics[width=.46\textwidth]{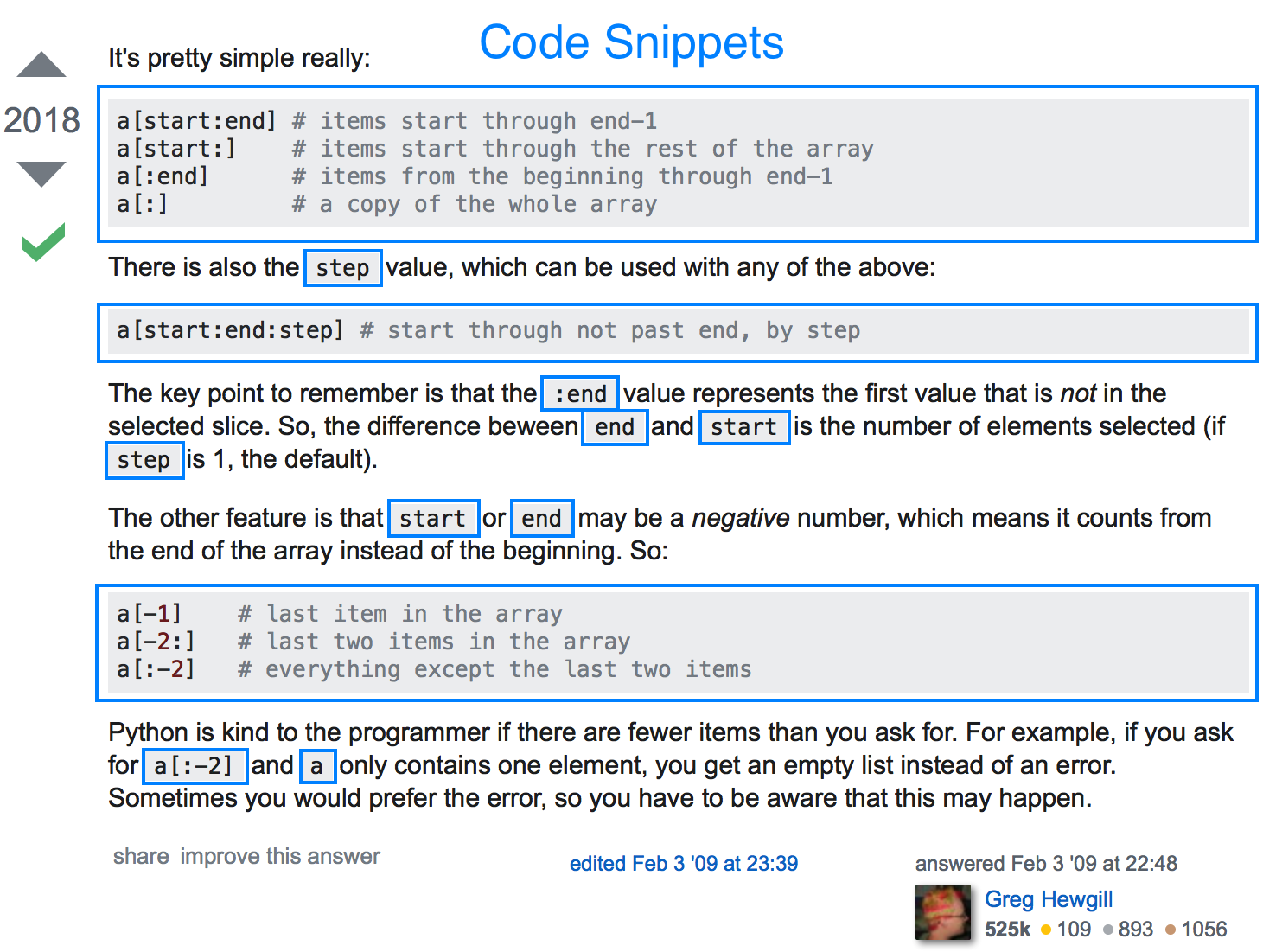}
\caption{Example Stack Overflow post showing code snippets.}
\label{fig:snippets}
\end{figure}

We model code segmentation as a classification problem, where given a sequence of characters, we must predict whether a character denotes the beginning of a new code segment. In order to generate training data suited to the task, we use Stack Overflow, a forum where users can ask questions, receive answers, and post code snippets on a wide range of computer programming topics. Because the posters are focused on answering a specific question, the code snippets are generally geared towards a single logical task. Fig. \ref{fig:snippets} shows an example response on Stack Overflow containing blocks the user has marked as code. We pull code snippets by searching for posts tagged with six different programming languages: C, C++, Java, Python, Javascript, and C\#.

\begin{figure}[t]
\includegraphics[width=.40\textwidth]{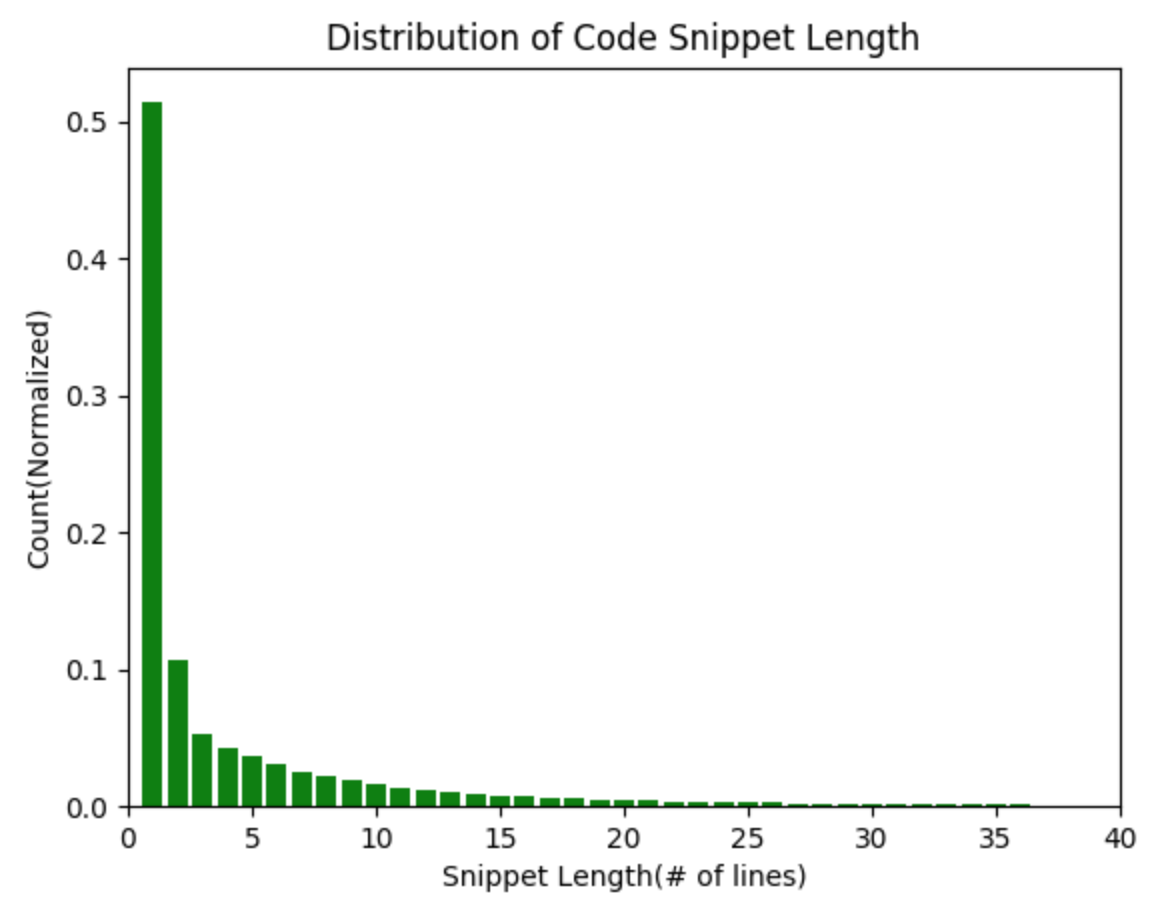}
\caption{Graph depicting the heavily skewed snippet length distribution on Stack Overflow.}
\label{fig:distribution}
\end{figure}

One problem with the data, however, is that the distribution of the number of lines per code snippet is heavily skewed. Fig. \ref{fig:distribution} shows that the majority of code snippets are only a few lines long. Using the entire data set could bias the model to predict segments every few lines because the model will have seen so many short code snippets, which is unlikely to be the case in real source code files. Using only very long snippets, however, would not leave much training data. Thus, we heuristically filter out all code snippets that are less than four lines long using the elbow method. 

\begin{figure}[t]
\includegraphics[width=\columnwidth]{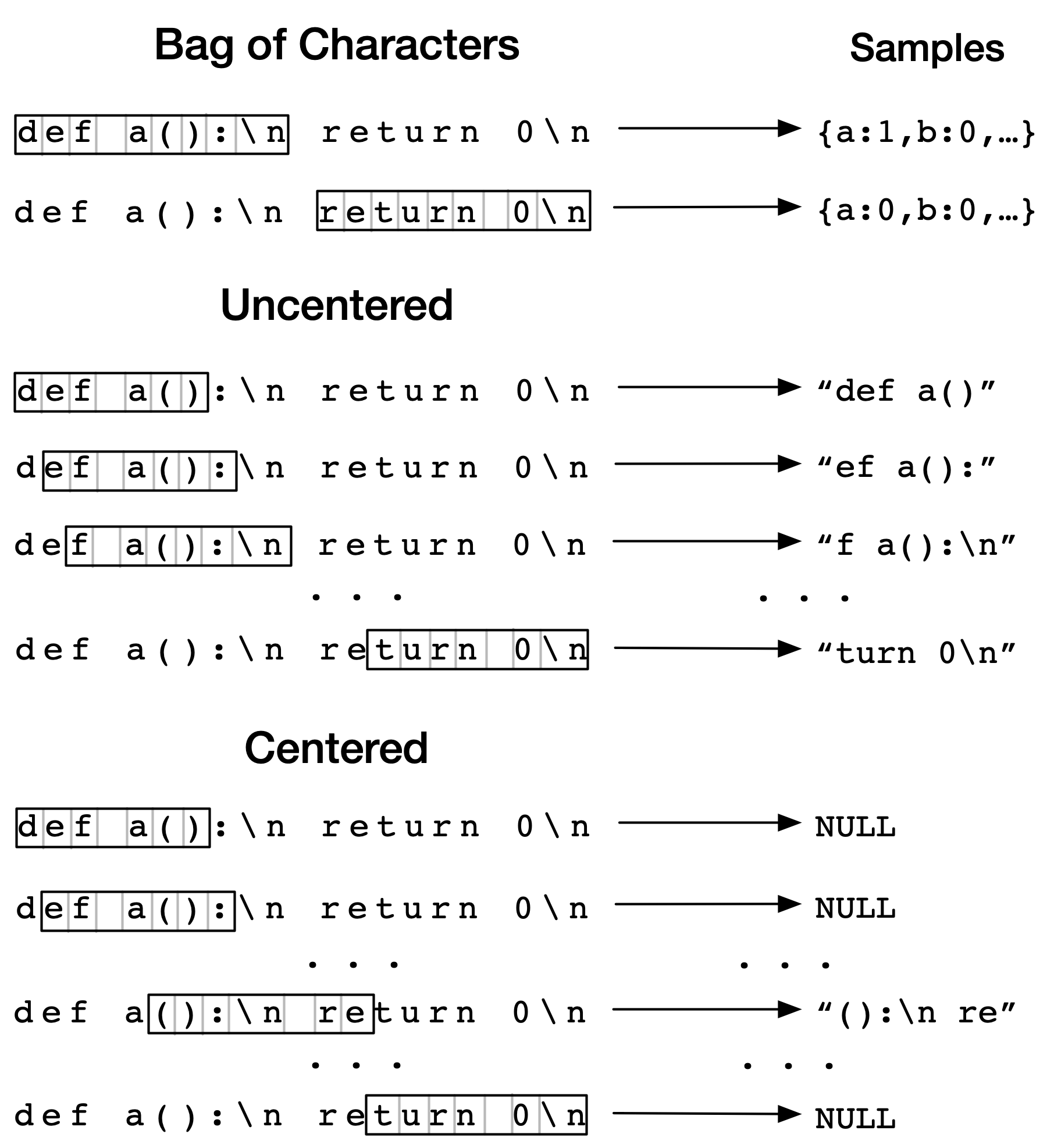}
\caption{The three data generation methods operating on the same piece of code. The window sizes are shortened for visual clarity. In the bag of characters method, each sample is the concatenation of seven bags of characters corresponding to seven consecutive lines. In the uncentered method, each sample is simply all the characters in the window. In the centered method, each sample is all the characters in a window where the middle character of the window is a newline. If a window in the centered method does not contain a newline in the center, no sample (``NULL'') is generated.}
\label{fig:datagen}
\end{figure}

After filtering, we generate segments by concatenating snippets with a newline character, thereby marking the beginning of a new segment. We refer to these as ``dividing newlines.'' It is important to note that not all newlines mark a new segment because a single code snippet may contain many newlines. After this process is complete, the result is essentially a giant block of concatenated code snippets. To obtain individual data points, we iterate through the block of snippets and generate data points using three methods: bag of characters, uncentered, and centered, shown in Fig. \ref{fig:datagen}.

In the bag of characters method, one data point is created by taking 7 lines from the block of snippets and counting the characters to create a ``bag of characters'' for each line. These bags of characters are then concatenated together to create a single training sample. If the middle newline (the fourth of seven) is a dividing newline, then the label for this data point is a 1. Otherwise, the label is a 0. This process is repeated by sliding the 7 line window forward by 1 line. We track the counts of 256 unique characters, corresponding to all the ASCII characters, with an average of about 29 characters per line.

In the uncentered method, we create a data point by selecting all the characters in a 100-character window. In this process, we assign one label for each character in the window, meaning one data point has 100 characters and 100 labels. If the character is a dividing newline, the corresponding label is a 1. This process is repeated by sliding the window by 100 characters. It is important to note that there is no guarantee a data point in this method will contain a dividing newline (or any newline at all). Even when the data point does contain a dividing newline, there is no guarantee that it will occur in the center of the input.

In the centered method, we create a data point by locating a newline and taking a window of 50 characters before and after that newline, for a total of 101 characters. If that newline is a dividing newline, then the label for this data point is a 1. Otherwise, the label is a 0. This process is repeated by centering the window on the next occurring newline. Whilst iterating through the block of snippets, any window that does not have a newline in the center will be ignored. It is important to note that other newlines, including dividing newlines, may occur in the window; however, the label is assigned based only on the center newline.

\section{\uppercase{Methodology}}
\label{sec:methodology}

We experiment with three models: a logistic regression model (to serve as a baseline), and two neural network architectures that utilize bidirectional LSTM layers. Every model is trained on seven training sets. Six of the training sets correspond to the different languages: C, C++, Java, Python, Javascript, and C\#. The last training set combines all of the languages in order to evaluate language agnosticism.

The logistic regression model utilizes the bag of characters data format for training. Each bag of characters, corresponding to one line of code, contains counts for 256 unique characters. The model takes seven bags of characters, for a total input size of 1,792. The output is a value from 0 to 1, representing the probability that the bag of characters corresponding to the fourth line (the middle line out of 7) contains a dividing newline (recall the labeling scheme from section \ref{sec:data}). 

The first neural network architecture uses the uncentered data format for training. This model has an input layer of length 100, one for each character in an uncentered data point. Each character is passed through a 20-dimensional character embedding, which converts a character to a 20-dimensional, real-valued vector. These embeddings are passed to a bidirectional LSTM layer of size 256. A bidirectional LSTM allows for past and future information to be used together, whereas a standard LSTM only considers past information. If a human were to segment source code, they would likely look ahead and use future information to piece together their decisions. The sequential information that the bidirectional LSTM learns is condensed using three time-distributed dense layers, sizes 150, 75, and 1, respectively. Since the layers are time distributed, the last layer has 100 total outputs (one for each time step). Each output is the probability that the character at that time step indicates the beginning of a new logical segment. The uncentered data format has 1 label for each character, so the 100 outputs and 100 labels are used to compute the loss. We use a batch size of 128, a dropout strength of 0.2 between each layer for regularization, and binary cross entropy as our loss function. 

The second neural network architecture uses the centered data format for training. In terms of model structure, it is nearly identical to the uncentered model except that the input size is 101 characters and that it has one additional layer. The last layer of the centered model is a dense layer that maps the 100 time distributed outputs (last layer of the uncentered model) to a single output. Because the centered data format guarantees a newline at the center of the input sequence, only a single output is required from the centered model. That single output and the label for the center newline are used to compute the loss.

\subsection{Evaluation}

Due to our data generation method, only a newline character can denote the beginning of a new code segment. As a result, we specifically measure the newline accuracy of our models. This distinction is critical because the logistic regression and centered models predict only on newlines, while the uncentered model outputs a prediction for every character, which would skew accuracy. The newline accuracy metric is the percentage of newlines that are classified correctly as either a dividing newline or non-dividing newline. 

We split the collection of code snippets into train/validation/test sets of 80\% / 10\% / 10\%. The training process is stopped when the model does not improve after 20 epochs. 

\section{RESULTS}
\label{sec:results}

In total, we train 21 different models: 18 single-language models and 3 multi-language models. For each of the three architectures (logistic regression, uncentered LSTM, centered LSTM), we train a model on each language individually, as well as a model on all the languages simultaneously. Table \ref{table:newlineacc} displays the newline accuracies for each model/language pairing.

\def\stoptable#1{%
   \hline\endtabular\par\vspace{\abovecaptionskip}%
   {\footnotesize #1}%
   \endminipage}
\begin{table*} 
\captionof{table}{Newline Accuracies of Each Model} \label{tab:title} 
\vspace{.15in}
\centering
\begin{tabular}{|c|l|l|l|l|l|l|}
  \hline
  \multirow{2}{*}{\textbf{Model}} 
          & \multicolumn{6}{|c|}{\textbf{Language}} \\             \cline{2-7}
  & \textit{\textbf{C}} & \textit{\textbf{C++}} & \textit{\textbf{Java}} & \textit{\textbf{Python}} & \textit{\textbf{Javascript}} & \textit{\textbf{C\#}} \\  \hline
  Single-Language Logistic Regression & 95.26 & 95.2 & 95.6 & 91.61 & 94.53 & 95.63  \\      \hline
  Single-Language Uncentered LSTM & \textbf{98.8} & \textbf{98.76} & \textbf{98.96} & \textbf{99.17} & \textbf{99.2} & \textbf{99.3}  \\     \hline
  Single-Language Centered LSTM & 97.35 & 98.35 & 98.84 & 97.82 & 98.79 & 98.87 \\ \Xhline{3\arrayrulewidth}
  Multi-Language Logistic Regression & 94.76 & 94.81 & 95.1 & 90.76 &  93.48 & 95.09 \\ \hline
  Multi-Language Uncentered LSTM & 98.33 & 98.56 & 98.75 & \textbf{97.98} &  98.65 & 98.86  \\ \hline
  Multi-Language Centered LSTM & \textbf{98.5} & \textbf{98.62} & \textbf{98.82} & 97.83 & \textbf{98.73} & \textbf{98.95} \\ \Xhline{3\arrayrulewidth}
  Percent Non-Dividing Newlines & 92.06 & 92.12 & 92.35 & 91.10 & 92.02 & 92.12 \\ \hline
\end{tabular}
\label{table:newlineacc}
\caption*{\textmd{Newline accuracy test results for every model and language. ``Single-Language" models are trained and tested on one language at a time. ``Multi-Language" models are trained on every language simultaneously and then tested on each language separately. ``Percent Non-Dividing Newlines'' is the newline accuracy if a model were to always predict non-dividing for every newline.}}

\end{table*}

The logistic regression models perform the worst, but are still significantly better than predicting a non-dividing newline every time (non-diving newlines are the most common). This is an expected result because the logistic regression models do not take into account character interactions like the more complex models. The neural network models show a significant improvement in performance over the logistic regression baselines. 

Although the single-language models generally perform slightly better than their multi-language counterparts, the difference in newline accuracy is relatively small. This is a very positive result because the multi-language models are able to discern logical segments across languages with little to no performance hit. 

Another interesting comparison is the difference in performance between the uncentered and centered models. In the single-language category, the uncentered models outperform the centered models across the board. On the other hand, in the multi-language category, the centered models are usually more effective. One possible reason for this is the difference in code context between uncentered and centered data points. In the uncentered data format, it is possible for a data point to contain no newlines whatsoever. This may help the model understand the difference in context when a newline is actually present. In the multi-language scenario, however, snippets with no newlines may have a fundamentally different pattern for each language in the data set, which may significantly complicate what the model needs to learn. The centered multi-language models only need to learn the differences between languages as it pertains to the context around a newline, reducing the cost of learning on multi-language data. Given enough data and time, it may be possible that the multi-language models would be able to utilize the uncentered snippets more effectively.

One other noteworthy observation is the impact of similarity across languages. In the multi-language scenario, Python is the most syntactically distinct language and consistently performs the worst. It is feasible that the models are able to transfer knowledge across languages, so similar languages may benefit from each other's data. The differences in results between languages may also speak to the quality of Stack Overflow snippets for those languages. It is possible that different programming languages attract different questions, topics, and code snippet qualities on Stack Overflow, which would ultimately influence the model's performance. 

\subsection{Testing on Source Code}
 
\begin{figure}[t]
\includegraphics[width=.47\textwidth]{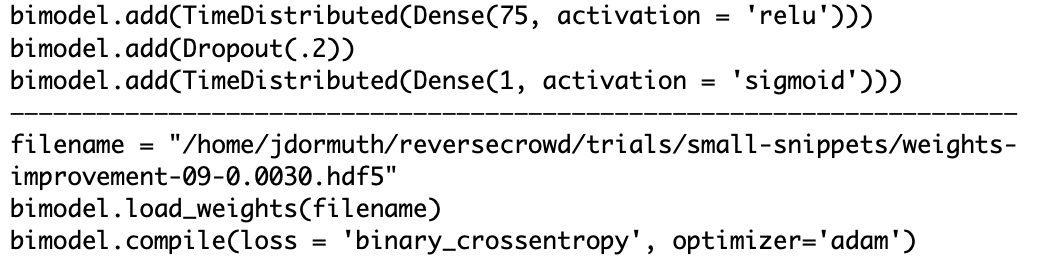}
\caption{The model recognizes that compiling and loading the weights of a Keras model is a different task from defining the layers of the model. The dashed line is the model's prediction of the segment location.}
\label{fig:divex1}
\end{figure}

\begin{figure}[t]
\includegraphics[width=.47\textwidth]{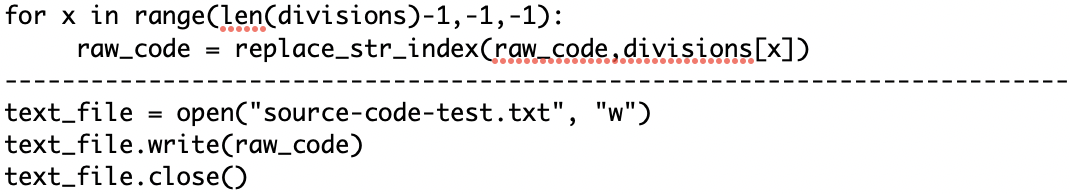}
\caption{The model distinguishes between string operations and writing to a file. The dashed line is the model's prediction of the segment location.}
\label{fig:divex2}
\end{figure}

In order to better understand the utility of these models, Fig. \ref{fig:divex1} and \ref{fig:divex2} showcase examples of the single-language deep learning model running on a Python source file.

In Fig. \ref{fig:divex1}, the model is able to recognize when the functionality of the code changes from defining a Keras model to loading the weights and compiling the model. Fig. \ref{fig:divex2} shows how the model is able to differentiate between a string operation in a for-loop and a new task of opening and writing to a text file.

\section{\uppercase{Challenges and Limitations}}
\label{sec:challenges}

One of the biggest challenges is the lack of ground truth logical segments. Because the Stack Overflow code snippets are simply concatenated together, the resulting data is not necessarily representative of real code. It is possible that two snippets concatenated together do not result in syntactically correct code. There is also no guarantee of standard formatting practices that one would expect to see in formal software projects. 

Additionally, the data generation method assumes that a code snippet represents a logical chunk of code. Although this is usually the case, there is no way to guarantee that all code snippets are segmented logically. It is possible that an individual snippet contains multiple logical tasks. It is also possible that two randomly selected snippets perform similar tasks; this introduces noise to the data set because they will still be labeled with a dividing newline. 

One possible way to address these challenges is to use in-line comments in source code files as logical division points. There is intended meaning behind the placement of comments in source code files, whereas the concatenation of Stack Overflow code snippets is randomized. This method could better reflect the properties of source code. 

% Lastly, augmenting our current data set to make it more natural and representative of real life source code could allow the model to be more robust and useful. A simple start to this task could be to randomly add spaces and newlines throughout the code snippets. 

% For example, concatenating two code snippets that both perform file read operations is not an optimal training sample. However, a sample that concatenates a code snippet that reads from a file and a code snippet that computes a mathematical formula calculation is much better since the topics are clearly logically different. 

\section{\uppercase{Conclusions and Future Work}}
\label{sec:conclusions}

We present a novel method to perform logical segmentation of source code. Using crowd-sourced data from Stack Overflow, we create a unique data set construction technique to approximate logical segments in source code. Drawing from the NLP domain, we develop deep neural network models utilizing bidirectional LSTMs that can predict on source code regardless of language or syntactical correctness. Lastly, we provide baselines and an appropriate metric to evaluate the performance of our models with regard to our data set construction. 

Although our method is the first success in language-agnostic logical code segmentation, there are a variety of potential architectural and parameter improvements. We could incorporate an attention mechanism into the LSTM, allowing the model to learn more specific features in the source code. Another avenue could be adjusting model parameters such as window size, network depth, and loss functions. For example, the loss function of the segmentation models could be modified to incorporate the error term of another task that uses segments as input. Finally, input representations that are larger than character-scale may improve the models without significant increases in model complexity; word or token embeddings in addition to the character embedding may be able to achieve this. 

\section*{ACKNOWLEDGEMENTS}

\noindent This project was sponsored by the Air Force Research Laboratory (AFRL) as part of the DARPA MUSE program. We would like to thank Robert Gove and Casey Haber for their valuable feedback, support, and contributions to figures. We also thank Banjo Obayomi for infrastructure support. 

\bibliographystyle{abbrv}
\bibliography{example}{}

%\vfill
%\bibliographystyle{apalike}
%{\small
%\bibliography{example}}

%\section*{\uppercase{Appendix}}
%\noindent If any, the appendix should appear directly after the
%references without numbering, and not on a new page. To do so please use the following command:
%\textit{$\backslash$section*\{APPENDIX\}}

%\vfill
\end{document}